\documentclass[11pt]{article}
\pdfoutput=1
\usepackage{jheppub}
\usepackage{subfigure,graphicx}

\newcommand\beq{\begin{eqnarray}}
\newcommand\eeq{\end{eqnarray}}
\newcommand\be{\begin{equation}}
\newcommand\ee{\end{equation}}

\title{Unparticles and Anomalous Dimensions in the Cuprates}

\preprint{\today}

\author[a]{Andreas Karch,}
\affiliation[a]{Department of Physics, University of Washington, Seattle, WA, 98195-1560, USA}
\author[b]{Kridsanaphong Limtragool,}
\affiliation[b]{Department of Physics and Institute for Condensed Matter Theory,
University of Illinois
1110 W. Green Street, Urbana, IL 61801, USA}
\author[b,1]{Philip W. Phillips\note{Guggenheim Fellow}}

\abstract{Motivated by the overwhelming evidence some type of quantum criticality underlies the power-law for the optical conductivity and $T-$linear resistivity in the cuprates, we demonstrate here how a scale-invariant or unparticle sector can lead to a unifying description of the observed scaling forms. We adopt the continuous mass formalism or multi band (flavor) formalism of the unparticle sector by letting various microscopic parameters be mass-dependent.   In particular, we show that an effective mass that varies with the flavor index as well as a running band edge and lifetime capture the AC and DC transport phenomenology of the cuprates.  A key consequence of the running mass is that the effective dynamical exponent can differ from the underlying bare critical exponent, thereby providing a mechanism for realizing the fractional values of the dynamical exponent required in a previous analysis\cite{ Hartnoll:2015sea}. We also predict  that regardless of the bare dynamical exponent, $z$, a non-zero anomalous dimension for the current is required. Physically, the anomalous dimension arises because the charge depends on the flavor, mass or energy.   The equivalent phenomenon in a $d+1$ gravitational construction is the running of the charge along the radial direction.  The nature of the superconducting instability in the presence of scale invariant stuff shows that the transition temperature is not necessarily a monotonic function of the pairing interaction.  }

\begin{document}
\maketitle


\section{Introduction}

Since the development of marginal Fermi liquid phenomenology\cite{mfl}, quantum criticality has been widely invoked to explain the observed power laws in both the DC\cite{qcrit1,qcrit2,qcrit3,qcrit4,qcrit5} and AC\cite{Marel2003,Anderson1997,ElAzrak1994,Schlesinger1990,Hwang2007,Basov2011} transport properties of the copper-oxide superconductors.  Because the underlying system is strongly correlated, a microscopic description of the degrees of freedom that are responsible for the quantum critical state is still lacking.  The difficulty in constructing even a phenomenological theory of criticality in the cuprates is evident from experimental observations of $T-$ linear resistivity  ($\rho$) and $\omega^{-2/3}$ scaling of the optical conductivity ($\sigma(\omega)$) for $\omega\gg T$.  The standard (no anomalous dimensions, no hyper scaling violation) implementation of single-parameter scaling places severe restrictions on the possible temperature\cite{chamon2004}, $\rho\propto T^{(2-d)/z}$ , and frequency\cite{wen1992scaling}, $\sigma(\omega)\propto \omega^{(d-2)/z}$ dependences. Here $d$ is the spatial dimension and $z$ the dynamical critical exponent.  In both of these expressions, the factor of $2$ in the exponent arises because the conductivity is determined by two derivatives of the action with respect to the vector potential whose scaling dimension is $d_A=1$.   As is evident, there is no accounting for both $T-$ linear resistivity and the $\omega^{-2/3}$  simultaneously by adjusting $d$ and $z$.   

While deviations from this can arise from the reduction of the effective dimensionality (hyper scaling violation with an exponent $\theta\ne 0$) through the presence of an additional length scale\cite{chamon2004} as in the presence of a Fermi surface and a non-integer dynamical exponent\cite{sachdev2015}, another distinct possibility is that the vector potential acquires\cite{Hartnoll:2015sea} an anomalous dimension, $\Phi$.  Within single parameter scaling, this modifies the scaling form of the conductivity (at $\theta=0$) to
\beq\label{daeq}
\sigma(\omega)\propto \omega^{(d-2d_A)/z},\quad d_A=1-\Phi.
\eeq
The desired power law of $-2/3$ requires that $2\Phi/z=-2/3$ which contradicts the Lorentz ratio result that $2\Phi/z=-1$\cite{Hartnoll:2015sea}.  Hence, within single-parameter scaling, there is no consistent scaling analysis of the transport observables even if an anomalous dimension is included in the current.  

Nonetheless, progress can be made by invoking the presence of a multi scale or unparticle\cite{Georgi2007} sector.
While there are numerous ways of formulating unparticles, the key insight stems from the fact that models with a large number $N$ of flavors can, in the large $N$ limit, create scale invariant theories with unusual scaling exponents. The scaling emerges from an interplay between the various flavors; even if the individual flavors themselves exhibit scaling with canonical exponents the behavior of the system as a whole is very different from the behavior of the individual flavors. Phillips and co-workers \cite{pp1,pp2} developed the continuous mass formulation of unparticles to address the general properties of the pseudo gap phase of the cuprates, whereas \cite{Karch:2015pha} referred to same construction as a ``multi-band model".   The latter emphasizes the fact that different flavors can be thought of as different bands, with $N$ standing for the number of bands in an energy region of interest. Scaling properties of thermodynamic quantities and DC transport in these multi-band theories have recently been studied in \cite{Karch:2015pha}, whereas dynamical processes including AC transport have been analyzed in \cite{Limtragool:2015nya}.

In \cite{Karch:2015pha} the band edge $M$ (that is the energy of the lower end of the band) and the charge $e$ of the carriers in the band where flavor dependent. In contrast \cite{Limtragool:2015nya} took the effective mass $m$ and the charge as well as the relaxation time $\tau$ to be flavor dependent. In a relativistic theory ``band edge" (that is rest mass) and effective mass are the same thing, but in a non-relativistic theory they are different concepts. For a simple free (parabolic) band the dispersion relation is given by
\beq
E = \frac{p^2}{2m} + M
\eeq
and the effective mass $m$ and the band edge $M$ are now completely separate parameters. While $M$ carries dimension of energy under the standard non-relativistic scaling, $m$ is actually dimensionless.
In this paper, we combine the analyses of \cite{Karch:2015pha,Limtragool:2015nya}, and study the most general multi-band model in which all four, $M$, $m$, $e$, and $\tau$ vary between flavors. This general construction has many appealing features:
\begin{itemize}
\item In \cite{Karch:2015pha}, where only $M$ and $e$ varied, the scaling dimension of the free energy as well the background electromagnetic fields in the multi-flavor model were already different from that of the individual flavors. This mismatch allowed for non-trivial exponents $\theta$ (the anomalous dimension for the energy density) and $\Phi$ (the anomalous dimension for the charge density). The dynamical critical exponent $z$ however was inherited from the individual flavors. In particular this meant that in order to realize the scaling exponents $\theta=0$, $\Phi=-2/3$ and $z=4/3$ that were shown in \cite{Hartnoll:2015sea} to give a successful fit to the DC phenomenology of the cuprates, one needed to start with $z=4/3$ for the individual flavors already, making it challenging to realize in solids. Allowing $m$ to vary as well allows to get dynamical critical exponent $z_*$ for the multi-flavor model unrelated to the underlying dynamical critical exponent $z$ of the individual flavors.
\item Letting in addition $\tau$ vary as in \cite{Limtragool:2015nya} one can combine the phenomenology of \cite{Hartnoll:2015sea} for DC transport with the successful prediction of \cite{Limtragool:2015nya} of the AC conductivity. Because of the relationship between $\tau$ and the breaking of diffemorphism in gravity contstructions, our work here provides restrictions on the possible radial-dependence of the emergent mass for the graviton.
\end{itemize}
We conclude our analysis with an analysis of the superconducting properties and show explicitly how a running mass affects $T_c$.

\section{DC properties}

\subsection{Review}

The starting point for the analysis of
\cite{Karch:2015pha} was a free energy density of the $n$-th individual flavors with band edge $M_n$, electric charge $e(M_n)$ and dynamical exponent $z$ as a function of temperature $T$ and background electro-magnetic fields $\mu$, $A_i$ given by:
\beq
\label{individualtwo}
\omega(\mu,T,A_i,M_n) = T^{\frac{d+z}{z}} f\left( \frac{e(M_n) \mu}{T}, \frac{e(M_n) A_i}{T^{1/z}},\frac{M_n}{T} \right ). \eeq
That is, in the action for the individual flavor $e(M_n)$, $\mu$ and $A_i$ only appear in the combination $e(M_n) \mu$ and $e(M_n) A_i$ and so any dependence on $e(M_n)$, $A_i$ and $\mu$ can only be in this product form. In addition, this free energy has a scale invariance with dynamical exponent $z$ as long as one treats the dimensionfull quantity $M_n$ as a spurion -- that is let it scale as an energy itself. For the multi-band model, whose free energy density is simply the sum over flavors, this implies in the large $N$ limit
\beq
\label{tobeintegrated}
\omega_{tot}(\mu,T,A_i) = T^{\frac{d+z}{z}} \, \int_0^\infty dM g(M) \, f\left( \frac{e(M) \mu}{T}, \frac{e(M) A_i}{T^{1/z}},\frac{M}{T} \right ).
\eeq
Here $g(M)$ is the density of levels in the sense that there are $g(M) dM$ flavors with band edge between $M$ and $M+dM$.
The full theory is scale invariant as long as $g(M)$ and $e(M)$ are given by power laws
\beq
g(M) = \frac{M^{a-1}}{ m_0^{a} }, \quad \quad e(M) = \frac{M^b}{ m_e^{b} }.
\eeq
$m_0$ and $m_e$ carry the dimension of energy. The power with which they appear in $g(M)$ and $e(M)$ respectively is determined by the fact that $e(M)$ is dimensionless whereas $g(M)$ has dimension of energy$^{-1}$.

The scaling properties of $\omega_{\rm tot}$ can be determined without any knowledge of the actual functional form of the free energy of the individual flavors by simply tracking how $m_0$ and $m_e$ appear in the final answer. $m_0$ only appears in $g(M)$, so it multiplies
$\omega_{\rm tot}$ as an overall prefactor $m_0^{-a}$. $m_e$ only enters where $e(M)$ appeared in the individual flavor free energy.   That is it will only occur in the combination $\mu/m_a^{b}$ and $A_i/m_e^{b}$, and neither $\mu$ nor $A_i$ will ever appear without $m_e$.  In addition, we know that $\omega_{\rm tot}$ will have to respect the underlying scale symmetry under which $m_e$ and $m_o$ scale as energies. These constraints together fix the functional form of $\omega_{\rm tot}$ to be given by
\beq
\label{integral}
\omega_{tot} = m_0^{-a} \, T^{\frac{d+z}{z} + a} \, \Omega \left ( \frac{\mu}{T} \left ( \frac{T}{m_e} \right )^{b} ,\frac{A_i}{T^{1/z}}  \left( \frac{T}{m_e}
\right )^{b} \right ) .
\eeq
The dependence of the total free energy on $T$, $\mu$ and $A_i$ corresponds to that of a scale invariant theory with exponents
\beq
\theta = - a z, \quad \quad \Phi=  b z, \quad \quad z_*=z.
\eeq

\subsection{Flavor dependent velocities}

The construction reviewed in the last subsection forces the multi-flavor theory to have the same dynamical critical exponent as the underlying single flavor theory. In order to separate these two exponents, it should be pretty clear that we need to make the speed of the various flavors different. If the individual flavor had $z=1$, this could be accomplished simply by making the dimensionless speed of light $v$ flavor dependent. By dimensional analysis, $\mu$ and $A_i$ differ in their dimensions by a factor of $v$ and we would postulate a free energy density per flavor of the form
\beq
\label{individualrel}
\omega(\mu,T,A_i,M) = T^{\frac{d+z}{z}} f\left( \frac{e(M) \mu}{T}, \frac{e(M) v(M) A_i}{T^{1/z}},\frac{M}{T} \right ). \eeq
For $z \neq 1$ the velocity itself has dimension $z-1$. The free energy density can still be taken of the form \eqref{individualrel}, but for $z \neq 1$ one has to keep in mind that in the end $v$ will be a function of $T$. In the $z=2$ case the best way to parametrize the flavor dependence of the velocity is as a flavor dependence of $m$. For $z=2$ the velocity of an excitation of momentum $p$ is given by $v(M) = p/m(M)$. The typical momentum in the system will be a non-trivial function of $\mu$ and $T$. The flavor dependence however will be completely encoded in the flavor dependence of $m$. With this, the free energy density per flavor takes the form
\beq
\label{individualnrel}
\omega(\mu,T,A_i,M_n) = T^{\frac{d+z}{z}} f\left( \frac{e(M) \mu}{T}, \frac{e(M) A_i}{T^{1/z} m(M)},\frac{M}{T} \right ). \eeq
For a scale invariant theory, we need in addition to $g(M)$ and $e(M)$ also $m(M)$ to take on a power law form,
\beq
m(M) = \frac{M^f}{m_m^f}.
\eeq
Integrating over flavors, we find that the scaling of the total free energy can once more to be worked out by tracking the appearance of $m_0$, $m_e$ and $m_m$. The only difference to the prior result is that this time $A_i$ will appear only in the combination $A_i m_m^f/m_e^{b}$. Correspondingly, the
total free energy density will be given by
\beq
\label{final}
\omega_{tot} = m_0^{-a} \, T^{\frac{d+z}{z} + a} \, \Omega \left ( \frac{\mu}{T} \left ( \frac{T}{m_e} \right )^{b} ,\frac{A_i}{T^{1/z}}\, T^{b-f} \, \frac{ m_m^f}{m_e^b} \right ) .
\eeq
Comparing this with the definition of the exponents $\Phi$, $\theta$ and $z$ according to which we should have
\beq
[\omega] = d+z_*-\theta, \quad \quad [\mu] = z_* - \Phi, \quad \quad [A_i] = 1 - \Phi
\eeq
the form \eqref{final} tells us that
\begin{eqnarray}
\nonumber z_*-bz_* &=& z_* - \Phi \\ \nonumber \frac{z_*}{z} - bz_* + fz_* &=& 1- \Phi\\
\frac{d+z}{z} + a &=& \frac{d+z_* - \theta}{z_*} \end{eqnarray}
which can easily be solved to yield
\beq z_* = \frac{z}{1+zf}, \quad \quad \Phi = b \frac{z}{1+zf}=b z_*, \quad \quad \theta = z \frac{df-a}{1+zf} = (fd- a) z_*. \eeq
We see explicitly that $z_\ast$ is no longer equal to the bare dynamical exponent, $z$.  Moreover, a flavor-dependent charge is the only mechanism by which the current can acquire an anomalous dimension, $\Phi$.

An explicit micsoscopic theory for unparticles that produces an anomalous dimension for the current is difficult to construct, however.   For example, if we define the total unparticle current,
\beq
j_U^{\mu}=\int g(m^2) j^\mu(m^2) dm^2,
\eeq
as a weighted sum over the currents for the m-dependent flavor fields,  it is clear that the scaling dimension of the total current and the current for the individual flavor fields will differ.    The total current is still a conserved quantity because by construction $\partial_{\mu} j^{\mu}=0$.  Hence, in principle, an anomalous dimension for the current and charge conservation are not necessarily contradictory.  However, in the standard implementation of unparticles from a quadratic action gauged with Wilson lines, no anomalous dimension survives.    This can be illustrated as follows.
A non-local quadratic action is in general given by
\beq
S = \int \frac{d^{d}p}{(2\pi)^d} \ \phi^\dagger(p)iG_U^{-1}(p)\phi(p),
\eeq
where $G_U(p)$ has a non-trivial scaling dimension of the form, $G_U(p) \sim (p^2)^{-\alpha}$.  Such an action can be generated by a non-canonical kinetic energy\cite{nontrad1}.
Introducing gauging through a Wilson line, we write the resulting action up to a term of order $A$ as
\begin{align}
S = &\int \frac{d^dp}{(2\pi)^d}\phi^\dagger(p)G^{-1}_U(p)\phi(p) + \int \frac{d^dpd^dq}{(2\pi)^{2d}}\phi^\dagger(p+q)\phi(p)A_{\mu}(q)g\Gamma^\mu(p,q) + O(A^2)
\end{align}
where
\begin{align}
g\Gamma^\mu(p,q) = & g(2p^\mu+q^\mu)\mathcal{F}(p,q) \\
\mathcal{F}(p,q) = &\frac{iG_U^{-1}(p+q) - iG_U^{-1}(p)}{(p+q)^2-p^2}.
\end{align}
The current is given by
\begin{align}
J^{\mu}(q) = \frac{(2\pi)^d}{V}\frac{\delta S}{\delta A_{\mu}(-q)} =  g\int \frac{d^dp}{(2\pi)^{d}}\phi^\dagger(p-q)\phi(p)\Gamma^\mu(p,-q) + O(A).
\end{align}
Using the convention $[p] = 1$, it follows that $[\phi(p)] = -(d+2\alpha)/2$ and $[\Gamma^\mu] = 2\alpha - 1$. From the scaling of these quantities, the scaling dimension of the current is just $[J(p)] = -1$.

However, the recent mapping of unparticles onto massive gravity does permit an anomalous dimension for the current.  The essential idea in this construction\cite{domokos} is to introduce a gauge transformation,
\beq
{\cal A}_\mu&\rightarrow& {\cal A}_\mu+\partial_\mu {\cal G},\nonumber\\
{\cal A}_y&\rightarrow& {\cal A}_y+\partial_y^{\alpha_A} {\cal G},
\eeq
which has a fractional derivative along one of the space-time coordinates, in this case the mass coordinate or the AdS radius.  This results in a field strength,
\beq
F_{\mu\nu}&=&2\partial_{[\mu}{\cal A}_{\nu]},\nonumber\\
F_{\mu y}&=&\partial_\mu{\cal A}_y-\partial_y^{\alpha_A}{\cal A}_\mu,
\eeq
and equations of motion
\beq
\partial_{\mu}\left(\sqrt{-g} F^{\mu\rho}\right)+\partial_y^{\alpha_A}\left(\sqrt{-g} F^{y\rho}\right)=0\nonumber\\
\partial_\lambda\left(\sqrt{-g} F^{\lambda y}\right)=0,
\eeq
which are satisfied only if all components of the vector potential acquire anomalous dimensions.  Consequently, the current also has an anomalous dimension.  Hence, massive gravity\cite{domokos} offers an explicit methodology of constructing unparticles from which the anomalous dimension is inherently manifest.  An open question to resolve with this methodology are the transport properties using the AdS dictionary.

\subsection{Fitting the cuprates}

We are now in a position to plug in numbers in order to reproduce the DC phenomenology of the cuprates. As in \cite{Hartnoll:2015sea}, we assume the copper oxide planes are $d=2$ dimensional. Also, in order to realize the system in terms of non-relativistic electrons we assume that the individual flavors/bands have standard dynamical exponent $z=2$. With these two assumptions the exponents $a$, $b$, and $f$ characterizing the flavor dependence of the parameters in our multi-band model are completely fixed by requiring that we reproduce the phenomenologically preferred values $z_* = 4/3$, $\Phi=-2/3$ and $\theta=0$. We find
\beq
f= 1/4, \quad \quad a=1/2, \quad \quad b=-1/2.
\eeq
If instead we chose $z=1$ for the individual flavors we get
\beq
f= - 1/4, \quad \quad a= -1/2, \quad \quad b=-1/2.
\eeq
Since the optical conductivity in the cuprates scales as $\omega^{-2/3}$, it might be tempting to assume that the relevant exponent here is $\Phi=-2/3$.  However,  we have pointed out previously that such a choice for $\Phi$ would lead to $\omega^{-1}$ assuming that $\theta=0$.  Hence, it is necessary to go beyond single-parameter scaling even if anomalous dimensions are allowed.

\section{AC properties}

In order to reproduce the universal $\omega^{-2/3}$ frequency dependence seen in the optical conductivity of the cuprates at large $\omega$, \cite{Limtragool:2015nya} introduced one more flavor dependent quantity: the lifetime $\tau(M)$. For the theory to have scaling properties, we once more have to postulate a power law form:
\beq \label{tau} \tau(M) = \frac{M^{C}}{m_{\tau}^{C+1}}. \eeq
Unlike $e$ and $m$, $\tau$ is dimensionfull (with dimension of inverse energy) and so the power law ansatz looks like that for $g$, rather than the ones for $e$ and $m$. Also note, that the power law we postulate for the lifetime has a qualitative different character from the power laws we demanded for $g(M)$, $e(M)$ and $m(M)$. The latter are properties of the distribution of the flavors. The physics of each individual flavor is insensitive to the dimensionfull constants $m_0$, $m_e$ and $m_m$ that characterize these distributions. What was important in the derivation of the previous sections was that the physics of the individual flavors only was dependent on a single dimensionfull quantity $M$. 
While in our toy examples $M$ was the band edge, all our derivations really required was that it was the only intrinsic dimensionfull scale on which the physics of a single flavor depends. What the power law \eqref{tau} demands is that the dependence of the lifetime of this individual flavor on this scale $M$ is a power law. So unlike the previous power laws, which where requirements on the distributions of flavors, \eqref{tau} actually makes assumptions about the dynamics of an individual flavor. In order to highlight this difference we used a capital letter $C$ for this exponent in order to highlight its qualitatively different nature. As a consequence, the ``new" scale $m_{\tau}$, has to be completely determined in terms of the external parameters like $T$, $A_i$ and $\mu$. This is different from $m_0$, $m_e$ and $m_m$, which appeared as new dimensionfull constants characterizing the flavor distribution.

Breaking of diffemorphism invariance in massive gravity constructions\cite{vegh2008, drgt2012} through the graviton acquiring a mass is the mechanism for momentum relaxation.   While energy is still conserved, momentum is not and the relaxation time $\tau$ explicitly enters\cite{davison2013} the equations of motion
\beq
\partial_t T^{it}=-\tau^{-1}T^{it}
\eeq
of the stress energy tensor.   The exponent $C$ in these scenarios is determined by the radial dependence of the graviton mass.

The conductivity of the individual flavor is given by a standard Drude form
\beq
\sigma(M) = \frac{e(M)^2 n(M)}{m(M)} \frac{\tau(M)}{1 - i \omega \tau(M)}.
\eeq
Here $n(M)$ is the particle number density associated with the $M$-th flavor.  It is related by $n=-\partial_{\mu} \omega$ to the single flavor free energy density from \eqref{individualnrel}. In order to connect to the results of \cite{Limtragool:2015nya} we need to postulate that this quantity obeys a power law as well:
\beq  \label{numberdensity} n(M) = \frac{M^A}{m_n^{A-d/z}}. \eeq
Like \eqref{tau}, the power law \eqref{numberdensity}, is a postulate on the dynamics of the individual flavors and the dimensionfull quantity $m_n$ can not correspond to a new scale but needs to be entirely determined in terms of $\mu$, $T$ and $A_i$.
With this, the total conductivity then is given as a sum over flavors similar to how we obtained the free energy in previous sections:
\beq
\label{drude}
\sigma = \int dM \, g(M) \frac{e^2(M) \tau(M) n(M)}{m(M)} \frac{1}{1- i \omega \tau(M)}.
\eeq
The conductivity that follows from this expression is, by construction, consistent with the scalings derived in the previous section as long as we keep in mind that the dimensionfull constants $m_{\tau}$ and $m_n$ do not correspond to new scales but are fixed in terms of $A_i$, $\mu$ and $T$. Postulating the Drude form in \eqref{drude} is an additional dynamical assumption on the individual flavors and allows to make predictions for the AC behavior.
A few additional comments are in order to compare with \cite{Limtragool:2015nya}. In this work, all our $M$ integrals always go from 0 to infinity. This way all quantities are genuinely scale covariant. \cite{Limtragool:2015nya} cuts off the mass integral at some upper mass $M_c$. In this case the resulting exponents are identical to the ones we find here, but scaling is only valid at large frequencies. While the latter may be more realistic when applied to real materials, for now we are interested in the intrinsic properties of the scale invariant theory defined by the $N\rightarrow \infty$ limit of our theory and keep the integral all the way to infinity. Maybe more importantly, \cite{Limtragool:2015nya} labels the flavors by $m$, not $M$. In fact, the band edge $M$ does not appear at all in the Drude formula for the conductivity and so can be ignored for this particular calculation (even though it does matter in the thermodynamic considerations of the last section). The easiest way to compare our formulas here to \cite{Limtragool:2015nya} is to set $f=1$, in which case $m \sim M$ and it doesn't matter whether we integrate over $m$ or $M$. Furthermore, \cite{Limtragool:2015nya} assumed constant $n$, that is $A=0$. Substituting the power law scalings for the $M$ dependent
quantities into the formula for $\sigma$ one finds
\beq
 \sigma(\omega)=  {\cal A} \int dM \, \frac{M^{a+2b+C-1-f+A}}{1 - i \omega \frac{M^{C}}{m_{\tau}^{C+1}}}
\eeq
where ${\cal A}$ collects all the dimensionfull quantities defining our power laws
\beq
{\cal A} = \frac{m_m^f}{m_0^{a} m_e^{2b} m_{\tau}^{C+1} m_n^{A-d/z}}.
\eeq
Changing variables to
\beq
x = \omega M^C/m_{\tau}^{C+1} \eeq
the integrals can easily be evaluated\footnote{One should note that for $C<0$ infinite $M$ maps to zero $x$ and
vice versa, so that an extra sign is introduced if we want to keep the $x$ integral from zero to infinity.} as in \cite{Limtragool:2015nya}
\beq
\sigma(\omega) = \frac{\cal{A}}{|C|} \frac{1}{ \omega} \left ( \frac{m_{\tau}}{ \omega}
\right )^{\frac{a+2b-f}{C}} \, \int_0^{\infty} dx \frac{x^{\frac{a+2b-f+A}{C}}}{1-ix} \sim \omega^{- \alpha}
e^{\frac{i \alpha \pi}{2}}
\eeq
with
\beq \alpha = \frac{a + 2b -f+A}{C}+1 .\eeq
Reassuringly this reduces to the expression in \cite{Limtragool:2015nya} for $f=1$ and $A=0$. To obtain the experimentally relevant value of $\alpha=2/3$ given the parameters $f=1/4$, $a=1/2$ and $b=-1/2$ or $f=-1/4$, $a=-1/2$ and $b=-1/2$ we found in the last section for $z=2$ and $z=1$ respectively we have to fix the one remaining exponents $C$ and $A$ to
$C= 9/4-3A$ or $C=15/4-3A$. These relations correspond to particular dynamical assumptions about the individual flavors.

\section{Superconductivity}

\subsection{Scaling Form of the Spectral Function}

The goal of this section is to determine what is different about the superconducting instability in a model system with a large number of flavors or a large number of bands.  In our previous work on superconducting instabilities in the presence of unparticles, we formulated the problem entirely in terms of the scaling dimension of the propagator\cite{pp1,pp2}. No explicit mention was made of the mass-dependence of the charge, density of states, or band edge and hence no immediate connection could be made with the results from the scaling analysis of the AC and DC properties.  It is this gap that we bridge in this section.

We consider a multi-band model with energy dispersion of the form
\beq \label{eq:dispersion}
\varepsilon_{\pm}(k,M) = \pm\gamma\frac{k^z}{m(M)} \pm M - \mu
\eeq
where $k$ is a magnitude of $d$-momentum, $\gamma$ is a positive constant, $M$ is the band off-set, $\mu(M)$ is a chemical potential, and $m(M)$ is a band mass. The positive sign and negative signs denote particle band and hole band, respectively.    We will work with convention that the dimension of the momentum is $[p] = 1$ and that of energy is $[\omega] = [T] = z$.  To take into account the flavor dependent velocities, the band mass is chosen to be $m(M) = \frac{M^f}{m_m^f}$. As before, the chemical potential $\mu$ is fixed across all bands. The bands with $M < \mu$ have finite particle density, whereas the bands with $M > \mu$ have vanishingly small particle density.

With this dispersion, the propagator of the band of mass $M$ is given by
\beq
G_{\pm}(\omega,k,M) = \frac{1}{\omega - \varepsilon_{\pm}(k,M)}.
\eeq
Approximating the unparticle field as an explicit sum over flavors\footnote{This construction leads to the correct form for the unparticle propagator but does not account for the unparticle field having lacking particle content.  Hence, it is strictly a construction for obtaining the unparticle propagator.}
\beq \label{eq:unparticle_field}
\psi_{U}(x) \equiv \sum_{n}F_{n}(\psi^{+}_{n}(x) + \psi^{-}_{n}(x)) = \int dM g(M)F(M) (\psi^{+}(x,M) + \psi^{-}(x,M))
\eeq
where the density of level $g(M) = \frac{M^{a-1}}{m_0^a}$ and F(M) is an M-dependent coefficient function leads to\cite{Deshpande2008,pp1} the unparticle propagator
\begin{eqnarray} \label{eq:propagator}
G(\omega,k) &=& \int\limits_{0}^{\infty} dM g^2(M)F^2(M) (G_{+}(\omega,k,M) + G_{-}(\omega,k,M)) \nonumber \\
&=& \int\limits_{0}^{\infty}dM g^2(M)F^2(M) (\frac{1}{\omega - \varepsilon_{+}(k,M)} + \frac{1}{\omega - \varepsilon_{-}(k,M)}.)
\end{eqnarray}
The choice of $F(M)$ depends on the system being studied. For now we let $F(M) = \frac{M^r}{m_F^r}$.
The corresponding spectral function is
\begin{eqnarray}  \label{eq:spectral_funciton}
A(\omega,k) &=& -\frac{1}{\pi}\mathrm{Im} \ G(\omega+i0^+,k) \nonumber \\
&=& \int\limits_{0}^{\infty}dM g^2(M)F^2(M)(\delta(\omega - \varepsilon_{+}(k,M)) + \delta(\omega - \varepsilon_{-}(k,M))) \nonumber \\
&\equiv& A_{+}(\omega,k) + A_{-}(\omega,k).
\end{eqnarray}
Here we define the spectral functions from the particle and hole bands as
\beq
A_{\pm}(\omega,k) \equiv \int\limits_{0}^{\infty}dM g^2(M)F^2(M)\delta(\omega - \varepsilon_{\pm}(k,M)).
\eeq
We first consider $A_{+}$. To integrate over the delta function, we look at the zero of the argument of the delta function,
\beq
\omega - \frac{\gamma m_m^f k^z}{M^f} - M + \mu = 0.
\eeq
or
\beq
M^{f+1} - (\omega+\mu)M^f + \gamma m_m^f k^z = 0.
\eeq
It is clear that when $\omega + \mu < 0$, $M$ has no positive real solutions. We introduce an ansatz, $M = (\omega+\mu) h(\frac{\gamma m_m^f k^z}{(\omega+\mu)^{f+1}})$. Substituting $M$ into the equation above, one obtains
\beq
h^{f+1}(x) - h^f(x) + x = 0 \label{eq:delta_zero}
\eeq
where $x = \frac{\gamma m_m^f k^z}{(\omega+\mu)^{f+1}}$. Suppose there exist positive real solutions to this equation, $h_i(x)$. In this case, the integral can be performed as
\begin{eqnarray}
A_{+}(\omega,k) &=& \int\limits_{0}^{\infty}dM \frac{M^{2(a-1)}M^{2r}}{m_0^{2a}m_F^{2r}} \sum\limits_{i} \frac{1}{|1-\gamma m^f_m f \frac{k^z}{(\omega+\mu)^f+1}h_i^{f+1}(\gamma m_m^f \frac{k^z}{(\omega+\mu)^{f+1}})|} \nonumber \\
&& \times \delta(M - (\omega+\mu) h_i(\gamma m_m^f \frac{k^z}{(\omega+\mu)^{f+1}})) \nonumber \\
&=& (\omega+\mu)^{2a+2r-2}\sum\limits_{i} \frac{m_0^{-2a}m_F^{-2r}h_i^{2a+2r-2}(\gamma m_m^f \frac{k^z}{(\omega+\mu)^{f+1}})}{|1-\gamma m^f_m f \frac{k^z}{(\omega+\mu)^{f+1}}h_i^{f+1}(\gamma m_m^f \frac{k^z}{(\omega+\mu)^{f+1}})|}.
\end{eqnarray}
$A_{+}(\omega,k)$ is zero when $\omega+\mu < 0$ or when Eq. \ref{eq:delta_zero} does not have any positive real solutions.
For the case of $A_{-}$, the equation which gives the zero of the delta function is
\beq
M^{f+1} + (\omega+\mu)M^f + \gamma m_m^f k^z = 0.
\eeq
As opposed to the the case of $A_+$, there are no positive real solutions for $\omega + \mu > 0$. We substitute in an ansatz,  $M = |\omega+\mu| h(\frac{\gamma m_m^f k^z}{|\omega+\mu|^{f+1}})$ and find the same equation as Eq. \ref{eq:delta_zero} with $x = \frac{\gamma m_m^f k^z}{|\omega+\mu|^{f+1}}$.  Integrating over the delta function, one finds
\beq
A_{-}(\omega,k) = |\omega+\mu|^{2a+2r-2}\sum\limits_{i} \frac{m_0^{-2a}m_F^{-2r}h_i^{2a+2r-2}(\gamma m_m^f \frac{k^z}{|\omega+\mu|^{f+1}})}{|1-\gamma m^f_m f \frac{k^z}{|\omega+\mu|^{f+1}}h_i^{f+1}(\gamma m_m^f \frac{k^z}{|\omega+\mu|^{f+1}})|}
\eeq
and $A_{-}(\omega,k)$ is zero when $\omega + \mu > 0$ or when Eq. \ref{eq:delta_zero} does not have any positive real solutions. From Eq. \ref{eq:spectral_funciton}, the total spectral function is
\begin{eqnarray}
A(\omega,k) &=& A_{+}(\omega,k) + A_{-}(\omega,k) \nonumber \\
&=& |\omega+\mu|^{2a+2r-2}\sum\limits_{i} \frac{m_0^{-2a}m_F^{-2r}h_i^{2a+2r-2}(\gamma m_m^f \frac{k^z}{|\omega+\mu|^{f+1}})}{|1-\gamma m^f_m f \frac{k^z}{|\omega+\mu|^{f+1}}h_i^{f+1}(\gamma m_m^f \frac{k^z}{|\omega+\mu|^{f+1}})|}.
\end{eqnarray}
The frequency $\omega + \mu > 0$ part of $A(\omega,k)$ comes from the particle band and the $\omega+\mu < 0$ part comes from the hole band.
This spectral function obeys the scaling form,
\beq
A(\omega,k) = |\omega+\mu|^{\alpha_A}f(\frac{k}{|\omega+\mu|^{\alpha_k}})
\eeq
with $\alpha_A = 2a+2r-2$ and $\alpha_k = \frac{f+1}{z}$.

As discussed in Ref. \cite{Balatsky1993}, the spectral function with such scaling form violates the f-sum rule. In the case of $\mu = 0$, we can see that the spectral sum
\beq
\int\limits_{-\infty}^{\infty} d\omega A(k,\omega)  = \Lambda^{\alpha_A}\int\limits_{-\infty}^{\infty} d\omega A(\Lambda^{-\alpha_k}k,\omega)
\eeq
is not, in general, equal to $1$. Here we scale $\omega \rightarrow \Lambda \omega$. It is then necessary to introduce a cutoff in the integral over $M$, namely, $W$. This cutoff is an energy scale of the system, for example the bandwidth.  Returning no to the integral over the delta functions in $A_{\pm}$, we find that beyond the frequency $|\omega+\mu|h(\gamma m_m^f \frac{k^z}{|\omega+\mu|^{f+1}}) < W$, the spectral function from the continuous mass formalism is zero. Note that in the limit of large $\omega$ and $f > -1$, $h(\gamma m_m^f \frac{k^z}{|\omega+\mu|^{f+1}}) = 1$. So the spectral function is cutoff at $\pm W -\mu$. The actual spectral function is the sum of the spectral function of the low energy theory - our multiband model and the spectral function of the high energy theory involving interband processes.

We still need to make sure that the integral in Eq. \ref{eq:propagator} converges. We avoided the evaluation of this integral directly by taking its imaginary part. However, if the integral does not converge, the unparticle propagator of Eq. \ref{eq:propagator} has no meaning. The integrand $\sim M^{\alpha_A - 1}$ as $M \rightarrow \infty$ and $\sim M^{\alpha_A+f}$ as $M \rightarrow 0$. Therefore, the integral converges when $-1-f<\alpha_A<0$ or $\frac{1-f}{2} < a+r < 1$. The divergence from the upper limit is relaxed because of the UV cutoff discussed in the previous paragraph. So the integral converges when $\alpha_A > -f-1$ or $a+r > \frac{1-f}{2}$.

\subsection{Superconducting Instability}

We now investigate a superconducting instability in a system of unparticles interacting with a featureless s-wave interaction of the form, $V(k-k') = \lambda w^*_k w_{k'}$ where $\lambda$ is the coupling constant and the occupation $w_k = 1$ when $0<\xi(k)<\omega_c$ and $w_k = 0$ otherwise.
The additional problem of  a superconducting instability between electrons interacting via an algebraic unparticle interaction has also been considered\cite{leblanc}.  Here $\xi(k)$ is some general function of unit energy.  The restriction $0<\xi<\omega_c$ means pairing can only occur between two unparticles, each with energy less than the cutoff energy, $\omega_c$. In the case of the BCS theory, $\xi(k)$ is a kinetic energy and $\omega_c$ is the Debye energy, $\omega_D$.  We can define the dimensionless coupling as $g \equiv \lambda N(0) (\mathrm{Volume})^{-1}$. $N(0)$ is the density of states of $\xi(k)$ within the range $0<\xi(k)<\omega_c$ and is assumed to be constant. This assumption is reasonable if $\omega_c \ll W$.

The condition for the superconducting instability is the divergence of the pair susceptibitliy. In the ladder approximation, such a condition is given by the equation,
\beq \label{eq:sup_instab}
1 = \lambda\sum\limits_{n,k}|w_k|^2 G(\omega_n,k)G(-\omega_n,k).
\eeq
Solving this equation, one will find the relationship between the coupling constant $\lambda$ and the transition temperature $T_c$. Following the procedure of Ref. \cite{pp1}, Eq. \ref{eq:sup_instab} can be rewritten as
\beq \label{eq:chi_pair}
1 = \frac{g}{2}\int\limits_{-W-\mu}^{W-\mu} dxdy \int\limits_{0}^{\omega_c}d\xi A(x,\xi)A(y,\xi)\frac{\tanh \frac{x}{2T}+\tanh \frac{y}{2T}}{x+y}.
\eeq
To understand more about the behavior of the solution to this equation, we calculate the beta function $\frac{dg}{d\ln T}$ \cite{Balatsky1993,pp1}. We scale out $\tilde{T} \equiv \frac{T}{W}$. The result is
\begin{eqnarray}
1 &=& \frac{g}{2}\tilde{T}^{2+2\alpha_A}\int\limits_{-W/\tilde{T}-\mu/\tilde{T}}^{W/\tilde{T}-\mu/\tilde{T}}dxdy \int\limits_{0/\tilde{T}}^{\omega_c/\tilde{T}} d\xi (x+\frac{\mu}{\tilde{T}})^{\alpha_A}(y+\frac{\mu}{\tilde{T}})^{\alpha_A}f(\frac{\xi}{x+\frac{\mu}{\tilde{T}}})f(\frac{\xi}{y+\frac{\mu}{\tilde{T}}})\frac{\tanh \frac{x}{2W}+\tanh \frac{y}{2W}}{x+y}. \nonumber \\
&=& \frac{g}{2}\tilde{T}^{2+2\alpha_A}\int\limits_{-W/\tilde{T}}^{W/\tilde{T}} dxdy \int\limits_{0}^{\omega_c/\tilde{T}} d\xi x^{\alpha_A}y^{\alpha_A}f(\frac{\xi}{x})f(\frac{\xi}{y})\frac{\tanh \frac{x-\frac{\mu}{\tilde{T}}}{2W}+\tanh \frac{y-\frac{\mu}{\tilde{T}}}{2W}}{x+y - \frac{2\mu}{\tilde{T}}}.
\end{eqnarray}
In the second line, we make a change of variables $x \rightarrow x' = x + \frac{\mu}{\tilde{T}}$ and $x \rightarrow x' = x + \frac{\mu}{\tilde{T}}$. We take logarithm, derivative with respect to $\tilde{T}$, and then change the variables and rescale the integrals back,
\begin{eqnarray} \label{eq:beta1}
\frac{dg}{d\ln T} &=& -2(1+\alpha_A)g + \frac{g^2}{2}\omega_c\int\limits_{-W-\mu}^{W-\mu} dxdy A(x,\omega_c)A(y,\omega_c) \frac{\tanh \frac{x}{2T} + \tanh \frac{y}{2T}}{x+y}  \nonumber \\
&& + \frac{g^2}{2}\mu\int\limits_{-W-\mu}^{W-\mu} dxdy\int\limits_{0}^{\omega_c}d\xi A(x,\xi)A(y,\xi) \bigg( 2\frac{\tanh \frac{x}{2T} + \tanh \frac{y}{2T}}{(x+y)^2} - \frac{1}{2T}\frac{\mathrm{sech}^2 \frac{x}{2T} + \mathrm{sech}^2 \frac{y}{2T}}{x+y}\bigg) \nonumber \\
&& + g^2W \int\limits_{-W-\mu}^{W-\mu}dy\int\limits_{0}^{\omega_c}d\xi A(y,\xi)A(W,\xi)\bigg(\frac{\tanh \frac{y}{2T}+\tanh \frac{W-\mu}{2T}}{y+W-\mu} + \frac{\tanh \frac{y}{2T}-\tanh \frac{W+\mu}{2T}}{y-(W+\mu)}\bigg). \nonumber \\
\end{eqnarray}
We use the fact that $W\gg\mu,\omega_c$ to simplify the expression in the last term. In the special case of $\mu = 0$, the third term drops out. We left with
\begin{eqnarray} \label{eq:beta2}
\frac{dg}{d\ln T} &=& -2(1+\alpha_A)g + \frac{g^2}{2}\omega_c\int\limits_{-W}^{W} dxdy A(x,\omega_c)A(y,\omega_c) \frac{\tanh \frac{x}{2T} + \tanh \frac{y}{2T}}{x+y} \nonumber \\
&& + g^2W \int\limits_{-W}^{W}dy\int\limits_{0}^{\omega_c}d\xi A(y,\xi)A(W,\xi)\bigg(\frac{\tanh \frac{y}{2T}+\tanh \frac{W}{2T}}{y+W} + \frac{\tanh \frac{y}{2T}-\tanh \frac{W}{2T}}{y-W}\bigg). \nonumber \\
\end{eqnarray}
The beta function we obtained here differs from the beta function in Ref. \cite{pp1} because  of the cutoff, W, in the spectral function. Eqs. \ref{eq:beta1} and \ref{eq:beta2} suggest that if the first term, the term of order $O(g)$, domintates in some regions of $(g,T)$, the behavior of  the solution to Eq. \ref{eq:chi_pair} is controlled by $\alpha_A = 2a + 2r - 1$. If $\alpha_A < -1$, $\frac{dg}{d\ln T} > 0$, so $T_c$ increases as $g$ increases. If $\alpha_A > -1$, $\frac{dg}{d\ln T} < 0$, so $T_c$ decreases as $g$ increases. However, it turns out that, the second and the third term in Eq. \ref{eq:beta2} are not much smaller than the first term. So the criterion above, based on a comparison of  $\alpha_A$ and $-1$, is not complete.

\begin{figure}
        \centering
        \subfigure[$\ \alpha_A = 0.5$ \label{fig:full_alpha0.5}]{\includegraphics[scale=0.3]{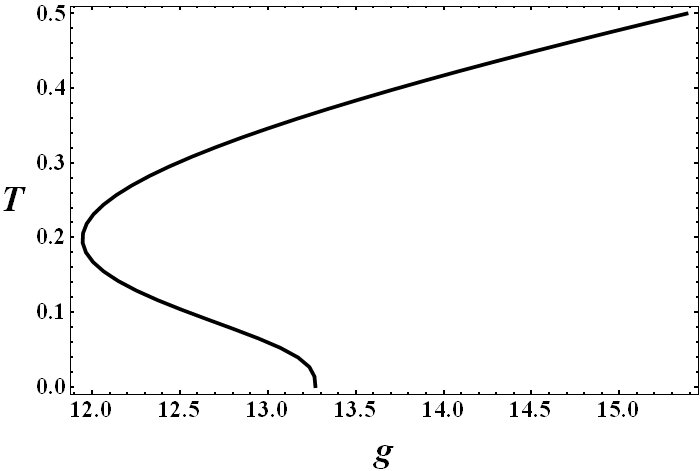}}
        \subfigure[$\ \alpha_A = 0$ \label{fig:full_alpha0}]{\includegraphics[scale=0.3]{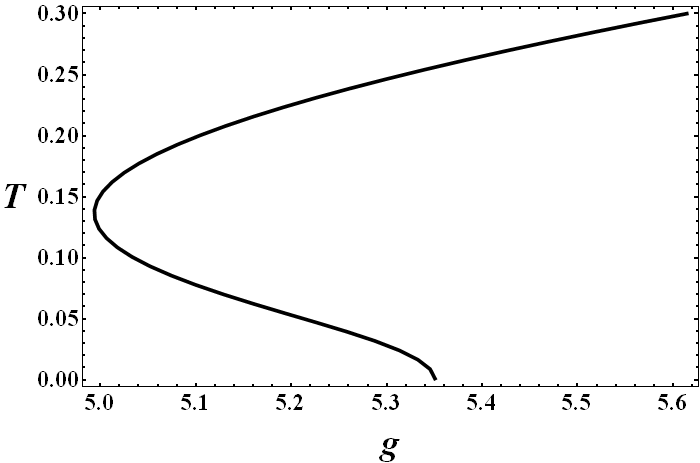}}
        \subfigure[$\ \alpha_A = -0.48$ \label{fig:full_alpha-0.48}]{\includegraphics[scale=0.3]{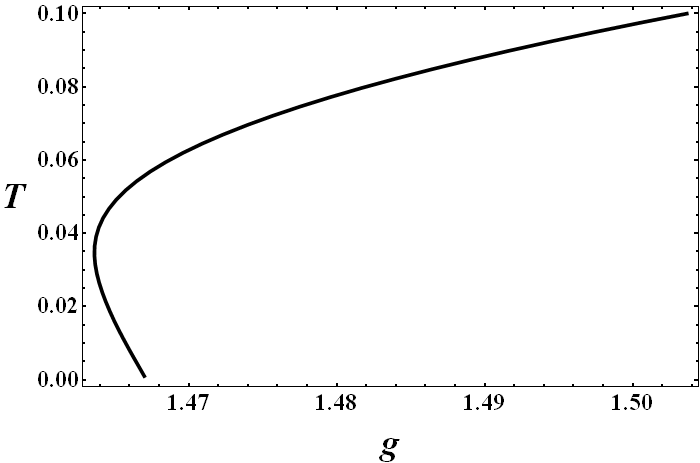}}
        \subfigure[$\ \alpha_A = -0.5$ \label{fig:full_alpha-0.5}]{\includegraphics[scale=0.3]{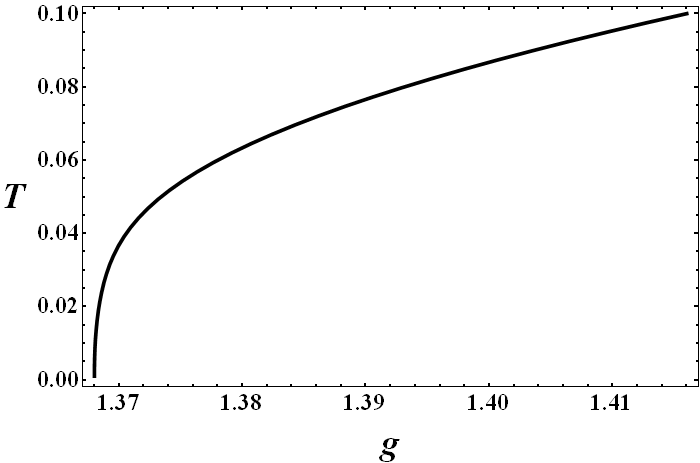}}
        \subfigure[$\ \alpha_A = -0.7$ \label{fig:full_alpha-0.7}]{\includegraphics[scale=0.3]{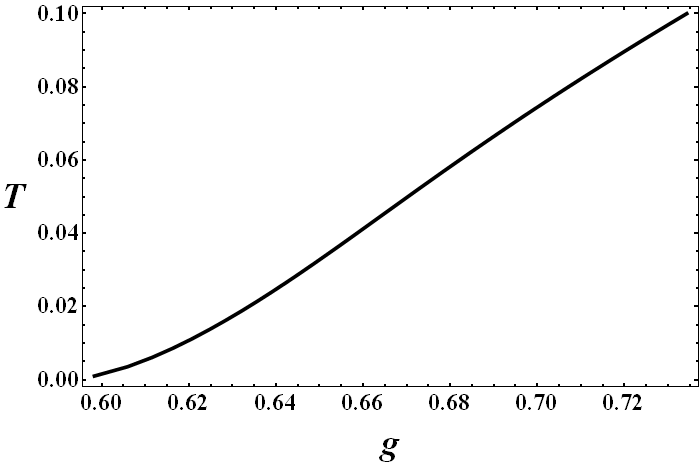}}
		\caption{Plots of $T_c$ vs. $g$ using the full spectral function $A = A_+ + A_-$. Note that for $\alpha_A < -0.5$, $\frac{dg}{d\ln T}$ is not negative for any $T_c$.} \label{fig:full}
\end{figure}

We demonstrate this behavior for the case of $z = 2$ and $f = 1$. We plot $T_c$ vs. $g$ using Eq. \ref{eq:chi_pair} for different values of $\alpha_A = 2a+2r-2$ in Fig. \ref{fig:full}. Here, we use $\omega_c = 0.1W$ and $\mu = 0$. We find that when $\alpha_A > -0.5$, $\frac{dg}{d\ln T}$ is negative at small value of $T_c$ and then becomes positive at larger value of $T_c$. When $\alpha_A$ is below $-0.5$, $\frac{dg}{d\ln T}$ is not negative at any $T_c$.

\begin{figure}
        \centering
        \subfigure[$\ \alpha_A = 0.5$ \label{fig:par_alpha0.5}]{\includegraphics[scale=0.3]{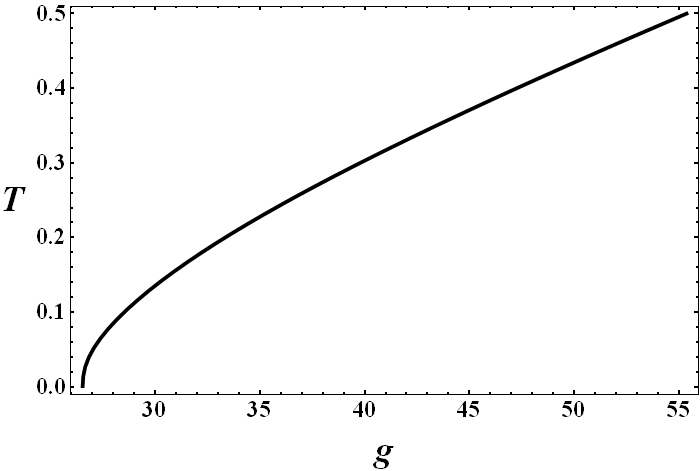}}
        \subfigure[$\ \alpha_A = 0$ \label{fig:par_alpha0}]{\includegraphics[scale=0.3]{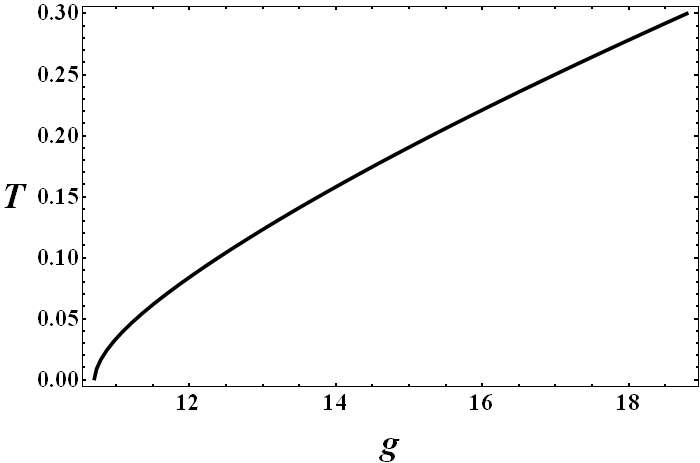}}
        \subfigure[$\ \alpha_A = -0.5$ \label{fig:par_alpha-0.5}]{\includegraphics[scale=0.3]{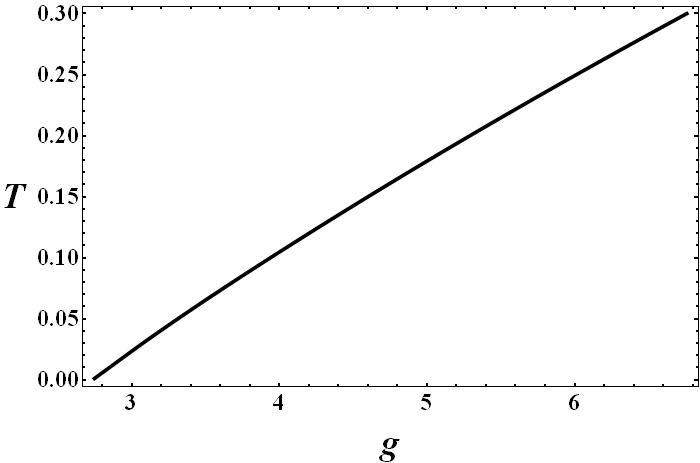}}
		\caption{Plots of $T_c$ vs. $g$ using only the particle spectral function $A_+$} \label{fig:par}
\end{figure}

The existence of the negative slope $\frac{dg}{d\ln T} < 0$ is strongly related to the shape of the spectral function.  By directly taking the derivative, one finds
\beq
\frac{dg}{d\ln T}= \frac{g^2}{4T}\int\limits_{-W-\mu}^{W-\mu}dxdy\int\limits_{0}^{\omega_c}d\xi A(x,\xi)A(y,\xi) \frac{x \ \mathrm{sech}^2(x/2T)+y \ \mathrm{sech}^2(y/2T)}{x+y}.
\eeq
As discussed in Ref. \cite{pp1}, with the appropriate form of the spectral function, the factor $\frac{x \ \mathrm{sech}^2x+y \ \mathrm{sech}^2y}{x+y}$ can be negative and outweigh the positive value in the $(x,y)$ space. We find that it is possible to obtain a negative $\frac{dg}{d\ln T}$ using the full spectral function $A = A_+ + A_-$. However, if we only use the spectral function from the particle band, $A_+$, $\frac{dg}{d\ln T}$ is always positive as shown in Fig. \ref{fig:par}.

\section{Closing Remarks}

Using the unparticle construction, we have shown how the current acquires an anomalous dimension from the an explicit flavor-dependent charge.  This is an explicit prediction of this work.  Further the unparticle construction leads to a power-law form for the scattering rate that varies with $z$.  We were able to fix the value of this power law, $c$ , by combining the AC and DC transport data.

The superconducting behavior of the multiband model is controlled by the types of bands that are being summed over and the scaling dimension of the resulting spectral function, $\alpha_A$. If only particle bands or only hole bands are used to construct the propagator, the transition temperature $T_c$ always increases with the coupling $g$, irrespective of the value of $\alpha_A$ (Fig. \ref{fig:par}). If both particle and hole bands are used, depending on the value of $\alpha_A$, it is possible to have a re-entrant behavior \cite{Balatsky1993}. That is, $T_c$ decreases with $g$ at small $T_c$, but increases with $g$ at larger $T_c$ (Figs. \ref{fig:full_alpha0.5}, \ref{fig:full_alpha0}, and \ref{fig:full_alpha-0.48}). This result suggests that  matter with scale invariant spectral functions needs to have the contribution from both particle ($\omega > -\mu$) and hole ($\omega < -\mu$) and the value of $\alpha_A$ must be large enough, in order for the re-entrant superconducting transition to appear. Since $\alpha_A = 2a + 2r - 2$, only the exponent $a$ (the exponent in the density of level $g(M)$) directly controls the behavior of $T_c$ vs. $g$. If the coefficient function $F(M)$ depends on $e(M)$, $\tau(M)$, $m(M)$, etc., other exponents such as $b$, $c$, and $f$ can affect behavior of superconductivity through the exponent $r$.

\section*{Acknowledgements}
P. Phillips thanks Sophia Domokos for a helpful conversation regarding massive gravity applied to unparticles.  We thank NSF DMR-1461952 for partial funding of this project.  KL is supported by the Department of Physics at the University of Illinois and a scholarship from the Ministry of Science and Technology, Royal Thai Government.
PP also acknowledges partial support from the Center for Emergent Superconductivity, a DOE Energy Frontier Research Center, Grant No. DE-AC0298CH1088 and the J. S. Guggenheim Foundation.

\providecommand{\href}[2]{#2}\begingroup\raggedright\endgroup

\end{document}